\documentclass[a4paper]{article}
\usepackage{tipa}

\usepackage{INTERSPEECH2021}

\usepackage{balance}

\usepackage[activate]{microtype}
\newcommand{\eg}{e.\,g.,\ }
\newcommand{\ie}{i.\,e.,\ }

\title{The EIHW-GLAM Deep Attentive Multi-model Fusion System \\for Cough-based COVID-19 Recognition in the DiCOVA 2021 Challenge}
\name{Zhao Ren$^{1}$, Yi Chang$^2$, Bj{\"o}rn W.\ Schuller$^{1,2}$}
\address{$^1$Chair of Embedded Intelligence for Health Care and Wellbeing, University of Augsburg, Germany\\
$^2$GLAM –- Group on Language, Audio, \& Music, Imperial College London, UK}
\email{zhao.ren@informatik.uni-augsburg.de, y.chang20@imperial.ac.uk}

\begin{document}
\maketitle
\begin{abstract}
Aiming to automatically detect COVID-19 from cough sounds, we propose a deep attentive multi-model fusion system evaluated on the Track-1 dataset of the DiCOVA 2021 challenge. Three kinds of representations are extracted, including hand-crafted features, 
image-from-audio-based deep representations, and audio-based deep representations. Afterwards, the best models on the three types of features are fused at both the feature level and the decision level. The experimental results demonstrate that the proposed attention-based fusion at the feature level achieves the best performance (AUC: $77.96\,\%$) on the test set, resulting in an $8.05\,\%$ improvement over the official baseline.

\end{abstract}


\section{System Description}
To describe our system, the methodology will be briefly introduced in Section~\ref{sec:method}. The pre-processing procedure of the audio signals will be then explained in~\ref{sec:preprocessing}. Afterwards, the methods of single-model feature extraction and classification will be described in Section~\ref{secsub:single-model}, and the fusion methods of multiple models will be given in Section~\ref{secsub:multi-model}. Finally, the experimental results will be shown and analysed in Section~\ref{sec:results}.

\subsection{Methodology Overview}
\label{sec:method}
Our proposed system consists of three major types of representations as the input: i) hand crafted features, ii) high-level image-from-audio-based representations extracted by pre-trained models from natural image datasets, and iii) high-level audio-based representations extracted by pre-trained models from audio datasets. The extracted features are then classified via training a feed-forward deep neural network (DNN) on the hand-crafted features or fine-tuning the transferred models on the high-level representations. The outputs of all classifiers are the probabilities of audio samples being COVID-19 positive. Finally, the multiple classifiers are assembled using a feature-/decision-level fusion. 

\subsection{Pre-processing}
\label{sec:preprocessing}
\begin{figure}[t]
\centering
\begin{minipage}[b]{.495\linewidth}
  \centering
  \centerline{\includegraphics[width=\linewidth]{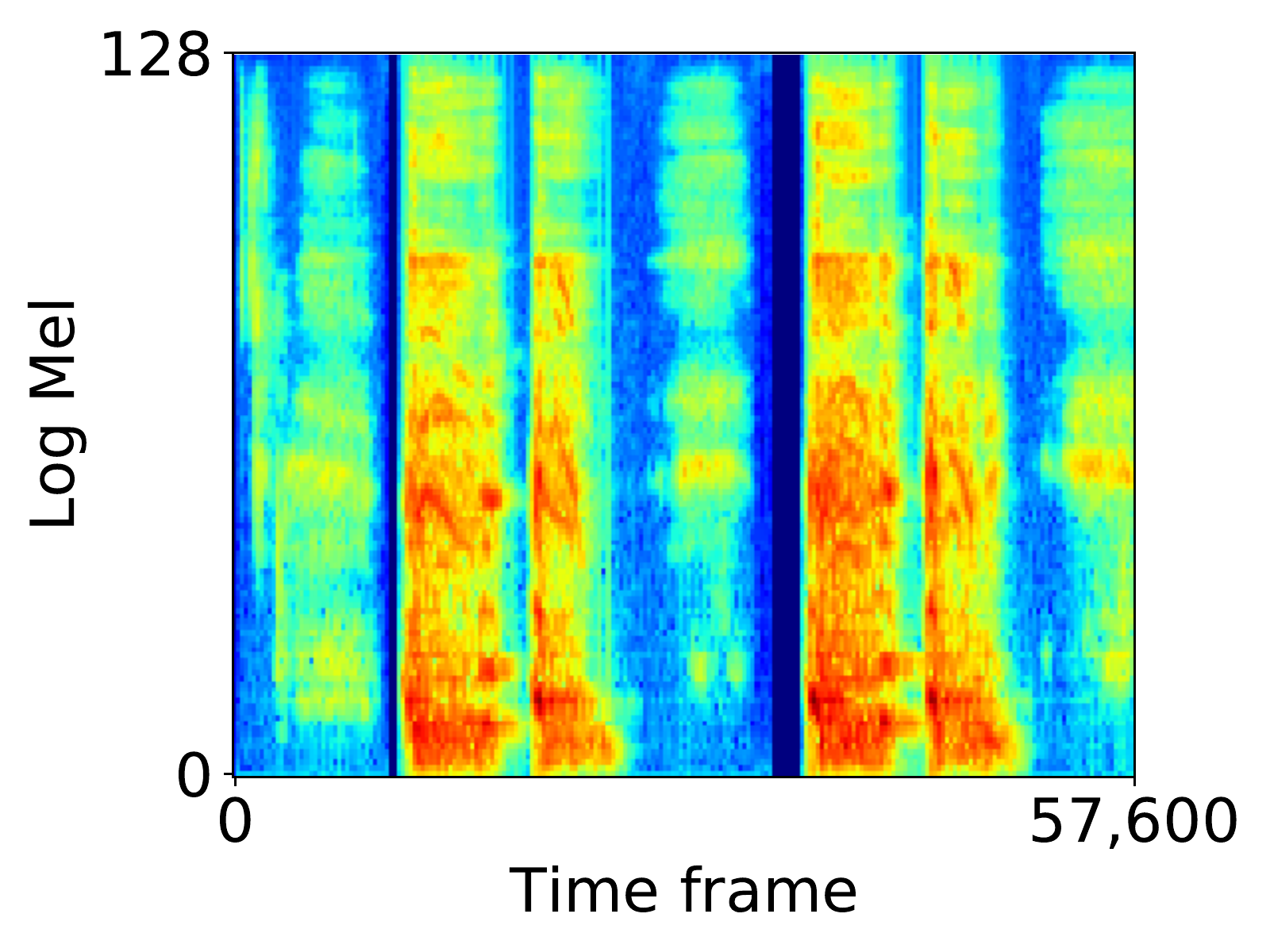}}
  \centerline{(a) Negative}\medskip
\end{minipage}
\begin{minipage}[b]{.495\linewidth}
  \centering
  \centerline{\includegraphics[width=\linewidth]{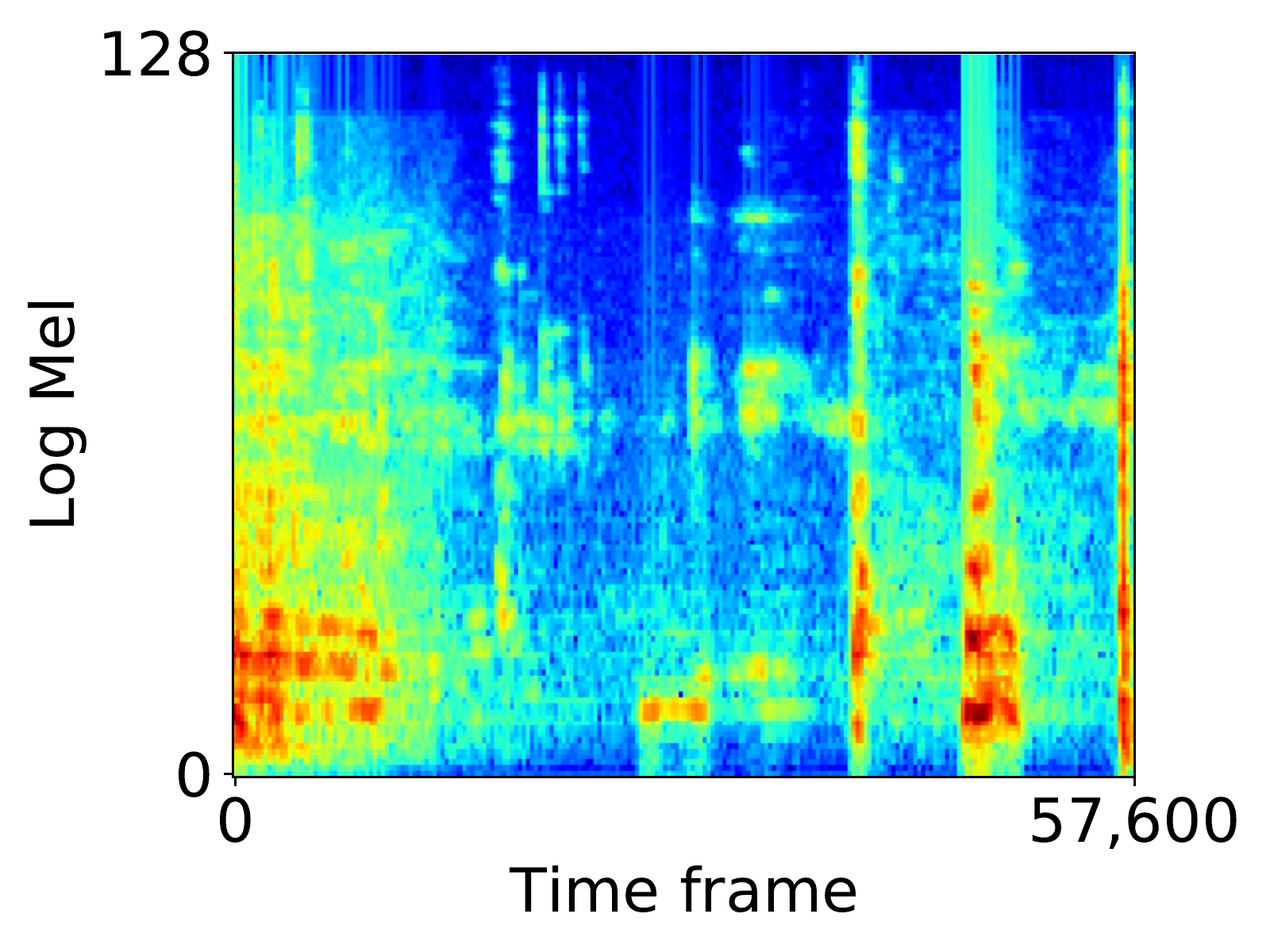}}
  \centerline{(b) Positive}\medskip
\end{minipage}
\caption{Two examples of the extracted log mel spectrograms from the Track-1 dataset of the DiCOVA 2021 challenge.}
\label{fig:logmel}
\end{figure}

In our experiments, the officially provided audio waves with a sampling rate of $44.1$\,kHz in the Track-1 of the DiCOVA challenge 2021~\cite{muguli2021dicova} are resampled into $16$\,kHz. To train a deep learning model, all of the audio signals are cut into smaller segments with a time length of $57,600$ frames. Notably, the log Mel spectrograms (cf. Figure~\ref{fig:logmel}) with a time length of $224$ frames are generated using a window size of $512$ and an overlap of $256$. 
For each audio sample, the probabilities of all its segments are averaged as the final prediction.
Further, to make the input size consistent with the hyperparameters in the pre-trained image-based and audio-based models for transfer learning (cf.\ Section~\ref{secsub:transfer_feature}), the number of Mel bins are respectively set to $64$ and $128$ for the audio-based models and the image-based models.

Before feeding the extracted features and log Mel spectrograms into deep learning models, mixup~\cite{kong2020panns,zhang2017mixup} is employed to augment the training data in our study. Specifically, the augmented data can be represented by $x=\alpha x_1 +(1-\alpha)x_2$ and $y=\alpha y_1 +(1-\alpha)y_2$, where $(x_1, y_1)$ and $(x_2, y_2)$ are two examples drawn from the original training data, and $\alpha$ is sampled from a Beta distribution. Next, both the original data and mixup-augmented data are processed by deep learning models in a training procedure.

\subsection{Single-model Feature Extraction and Classification}
\label{secsub:single-model}
The extraction and classification of the hand-crafted features and transfer-learning-based representations will be introduced in Section~\ref{secsub:handcrafted_feature} and Section~\ref{secsub:transfer_feature}, respectively.

\subsubsection{Hand-crafted Features}
\label{secsub:handcrafted_feature}
Hand-crafted features have been widely used and achieved good performances in audio/speech classification, such as heart sound classification~\cite{schuller2018compare}, cold speech detection~\cite{deb2017analysis}, etc. In this regard, three feature sets are used in our study: a  $2,600$-dimensional log Mel feature set, a $1,400$-dimensional Mel Frequency Cepstral Coefficients (MFCC) feature set, and a $6,373$-dimensional \textsc{Computational Paralinguistics ChallengE (ComParE)} 2016 feature set~\cite{schuller2018compare}. In the log Mel and the MFCC feature sets, $26$ Mel bins and $14$ coefficients are respectively calculated as the low-level descriptors (LLDs), and $100$ functionals are then applied to these LLDs. All of these features are extracted by the open source openSMILE toolbox~\cite{eyben2010opensmile}.

To predict the COVID-19 class (\textit{negative}/\textit{positive}), the hand-crafted features are fed into a DNN model, which contains three linear layers with the number of output neurons $1,024$, $256$, and~$1$, respectively.


\begin{table*}[t]
    \centering
    \caption{Performances of our proposed approaches evaluated on the Track-1 database of DiCOVA challenge 2021 for COVID-19 detection. The performances on the validation sets are the average results over the five folds, followed by the standard deviations (std).}
    \begin{tabular}{l p{1.8cm}p{1.8cm} p{1.8cm} p{1.8cm} p{1.8cm} p{1.8cm}}
    \toprule
    & \multicolumn{3}{c}{Validation} & \multicolumn{3}{c}{Test} \\ 
    \cmidrule(lr){2-4} \cmidrule(lr){5-7}
    [\%] & Sensitivity~(std) & Specificity~(std) & AUC~(std) & Sensitivity & Specificity & AUC \\
    \hline
    \multicolumn{7}{c}{\textit{Hand-crafted Features + DNN}} \\
    \hline
    Log Mel &\;  80.00 (0.00) &  33.16 (8.49) &  60.67 (6.15) &80.49 & 27.60 & 67.41 \\
    MFCC &\;  80.00 (0.00) &  30.88 (7.63) &  61.16 (3.65) &80.49 & 23.44& 55.63 \\
    \textsc{ComParE} &\;  80.00 (0.00) &  33.58 (8.08) &  65.50 (1.96) & 80.49 & 40.10 & 66.90 \\ 
    \hline
    \multicolumn{7}{c}{\textit{Transfer Learning from ImageNet}} \\
    \hline
    VGG11 &\;  80.00 (0.00) &  32.23 (2.73) &  63.09 (2.55) &80.49 &42.71 &65.69\\
    ResNet34 &\;  80.00 (0.00) &  20.41 (3.83) &  51.69 (3.80) &80.49 &38.02 &58.75 \\
    \hline
    \multicolumn{7}{c}{\textit{Transfer Learning from AudioSet}} \\
    \hline
    CNN14\_16k &\;  80.00 (0.00) &  42.28 (5.49) &  67.82 (3.29) &80.49 & 44.27 &67.77\\
    ResNet38 &\;  80.00 (0.00) &  39.27 (14.81) &  68.36 (4.54) &80.49 &45.31 & 65.66 \\
    \hline
    \multicolumn{7}{c}{\textit{Multi-model Fusion}} \\
    \hline
    Feature-Max & 100.00 (0.00) & 00.00 (0.00) & 63.73 (3.94) &80.49 &28.12 &65.56 \\
    Feature-Avg &\; 80.00 (0.00) & 35.54 (8.59) & 68.51 (2.09) &80.49 &61.46 &73.72 \\
    Feature-Attention  &\; 80.00 (0.00) & 44.56 (6.29) & 70.56 (3.01) &\textbf{80.49} &\textbf{59.38} &\textbf{77.96} \\
    \\
    Decision-Max &\;  80.00 (0.00) &  31.40 (8.20) & 66.66 (2.25) &80.49 &38.02 & 66.44\\
    Decision-Avg &\;  80.00 (0.00) &  35.03 (9.67) &  68.57 (2.98) &80.49 &55.73 &73.25 \\
    Decision-Attention &\; 80.00 (0.00) & 39.17 (9.41) &68.21 (3.92) &\textbf{80.49} &\textbf{59.38} &\textbf{77.36} \\    
    \bottomrule
    \end{tabular}
    \label{tab:result}
\end{table*}

\subsubsection{Transfer-learning-based Representations}
\label{secsub:transfer_feature}
To tackle small-scale datasets using deep learning topologies, transfer learning~\cite{pan2009survey} has shown potential of extracting highly abstract representations with pre-trained models learnt from large-scale datasets. Both image-based and audio-based pre-trained models are utilised to extract features from the Track-1 dataset of the DiCOVA challenge 2021.

\textit{Image-based models.}
A large number of deep learning models have proven to be effective in processing large-scale natural image datasets, such as ImageNet~\cite{deng2009imagenet}. In this work, we choose VGG11~\cite{simonyan2014very} and ResNet34~\cite{he2016deep} to extract high-level representations from log Mel spectrograms. Particularly, the final two linear layers of VGG11 are replaced with three new linear layers, which output the number of neurons $1,024$, $256$, and $1$. As only one linear layer is used in ResNet34 as the final layer, it is replaced with three new linear layers as those in VGG11, and updated along with the final convolutional block.
    
\textit{Audio-based models.}
While transferring pre-trained image-based models to audio classification tasks, the data difference between images and audio waves might be a potential bottleneck of improving the performance~\cite{koike2020audio}. Therefore, deep learning models trained on large-scale audio databases (\eg AudioSet~\cite{gemmeke2017audio}) are applied to extract representations from the log Mel spectrograms. We initialise our models with the parameters of the pre-trained CNN14\_16k and ResNet38~\cite{kong2020panns}, which have shown state-of-the-art performances on AudioSet. The first linear layer in both models is replaced with a trainable linear layer which outputs the number of neurons $1,024$, and followed by two trainable linear layers with the number of output neurons $256$, and~$1$.

\subsection{Multi-model Fusion}
\label{secsub:multi-model}
With the trained models on the hand-crafted features and the transfer-learning-based representations, a set of fusion approaches are employed to improve the performance. Especially, an attention mechanism~\cite{ren2018attention,ren2020caanet} is proposed in our study to fuse the multiple models. Next, the feature-level and decision-level fusion methods will be introduced in Section~\ref{secsub:featurefusion} and Section~\ref{secsub:decisionfusion}. 

\subsubsection{Feature-level Fusion}
\label{secsub:featurefusion}
We apply feature-level fusion methods at the output of the second linear layer in each best model of the three methods in Section~\ref{secsub:single-model} (\ie a DNN model on hand-crafted features, an image-based model, and an audio-based model), including max fusion, average fusion, and attention-based fusion. 

\textit{Max fusion} selects the maximum values from the learnt three representations, and outputs a vector with a length of $256$. The feature vector is finally processed by a linear layer, which outputs a probability value for each audio segment.

\textit{Average fusion} calculates the average vector of the three representations, and outputs a $256$ dimensional vector, which is then fed into a linear layer.

\textit{Attention-based fusion} aims to calculate the contribution of each unit in the three representations. An additional one-dimensional convolutional layer with a kernel size of $1$ and an output channel number of $256$, followed by a sigmoid function, is applied to the representations. Afterwards, the normalised output of the convolutional layer is multiplied with the representations. The summed results of the multiplication are finally fed into a linear layer to calculate the probability. 

\subsubsection{Decision-level Fusion}
\label{secsub:decisionfusion}
Apart from the feature-level fusion, decision-level fusion works on the output activation of the final linear layer. Similarly, the max, average, and attention-based decision-level fusion methods will be introduced as follows.

\textit{Max fusion} at the decision level selects the maximum probability.

\textit{Average fusion} computes the average probability of the obtained three probabilities for each audio segment.

\textit{Attention-based fusion} processes the output of the second linear layer with two one-dimensional convolutional layers, with a kernel size of $1$ and an output channel number of $1$. Next, one of the convolutional layers is followed by a sigmoid function and normalisation. Furthermore, the normalised output is multiplied with the output of the other convolutional layer, and the multiplication result is summed up to give the final probability.

\subsection{Experimental Results}
\label{sec:results}
\subsubsection{Database}
Our proposed approaches are evaluated on the dataset of Track-1 (cough sounds) of the DiCOVA 2021 challenge~\cite{muguli2021dicova}. The dataset is composed of $1,040$ audio files ($965$ non-COVID) recorded from COVID-19 positive and negative individuals. The majority of the individuals are COVID-negative male individuals with their age in $15-30$ years~\cite{muguli2021dicova}. All cough sound recordings are sampled to $44.1$\,kHz and compressed as FLAC format. The dataset is split into five folds for cross validation, where each fold has a training and a validation sets. 
Additionally, the blind test set is composed of $233$ audio samples. 
The evaluation metrics of the results consist of the sensitivity, the specificity, and the Area Under the Curve (AUC)~\cite{muguli2021dicova,fan2006understanding}.


\subsubsection{Experimental Setup}
During training, the parameters are updated with $30$ epochs and iterated with a batch size of $16$. The `Adam' optimiser with a learning rate of $0.001$ is experientially applied during the training process. As for the loss function, \texttt{BCEWithLogitsLoss} is used with the weight set as ratio of the number of COVID-positive samples to the number of COVID-negative ones in the training data. Moreover, after every $10$ epochs, the learning rate decays by $0.1$ to stabilise the training procedure. To recognise COVID-19 on the test set, the whole dataset with $1,040$ samples contributes to training models.

\subsubsection{Results and Discussion}

The results evaluated on the validation and test sets are given in Table~\ref{tab:result}. Compared to the official baseline (average validation AUC: $68.81\,\%$, test AUC: $69.91\,\%$), the results of the three single-model methods are comparable or slightly lower on the validation and test sets. However, the multi-model fusion approaches achieve improvements over the single-model methods, and mostly outperform the baseline system. Especially, the proposed feature-level attention and decision-level attention fusion approaches perform better than both the max and average fusion methods. Finally, the feature-level attention 
achieves an AUC of $77.96\,\%$ on the test set, resulting in an improvement of the baseline. 

In future efforts, more data pre-processing and augmentation techniques, such as signal enhancement and Generative Adversarial Networks (GANs) \cite{8683799}, could be explored. Meanwhile, more databases of cough sounds will be considered to augment the training data. Furthermore, COVID-19-related databases with more modalities, \eg speech signals, will be employed to achieve multi-modal COVID-19 detection for improving the performance.




\section{Acknowledgement}
We express our deepest sorrow for those who left us due to COVID19; they are lives, not numbers. We further express our highest gratitude and respect to the clinicians and scientists, and anyone else helping to fight against COVID-19 and maintain our daily lives. 
This work was supported by the Horizon H2020 Marie Sk\l{}odowska-Curie Actions Initial Training Network European Training Network (MSCA-ITN-ETN) project under grant agreement No.\,766287 (TAPAS). 

\balance
\bibliographystyle{IEEEtran}

\bibliography{mybib}

\end{document}